\documentstyle[aps,prb,psfig]{revtex}
\begin{document}

\title{Specific heat and thermal conductivity in the vortex state of the
two-gap superconductor MgB$_2$}
\author{L. Tewordt and D. Fay}
\address{I. Institut f\"ur Theoretische Physik,
Universit\"at Hamburg, Jungiusstr. 9, 20355 Hamburg, 
Germany}
\date{\today}
\maketitle
\begin{abstract}
    The specific heat coefficient $\gamma_s(H)$ and the electronic thermal 
conductivity $\kappa_{es}(H)$ are calculated for Abrikosov's vortex lattice
by taking into account the effects of supercurrent flow and Andreev scattering. 
First we solve the gap equation for the entire range of magnetic fields. We
take into account vertex corrections due to impurity scattering calculated in the
Born approximation. The function $\gamma_s(H)/\gamma_n$ increases from
zero and becomes approximately linear above $H/H_{c2} \sim 0.1\,$. The
dependence on impurity scattering is substantially reduced by the vertex
corrections. The upward curvature of $\kappa_{es}(H)/\kappa_{en}\,$, which
is caused by decreasing Andreev scattering for increasing field, is reduced for
increasing impurity scattering. We also calculate the temperature dependence
of the scattering rates $1/\tau_{ps}(H)$ of a phonon and $1/\tau_{es}(H)$ of a 
quasiparticle due to quasiparticle and phonon scattering, respectively. At low
temperatures the ratio $\tau_{pn}/\tau_{ps}(H)$ increases rapidly to one as $H$
tends to $H_{c2}$ which yields a rapid drop in the phononic thermal 
conductivity $\kappa_{ph}$. Our results are in qualitative agreement
with the experiments on the two-gap superconductor MgB$_2\,$.
\end{abstract}
\vspace{0.5in}
\section{INTRODUCTION}

          Although superconductivity in MgB$_2$ \cite{Nagamatsu} has been
established as conventional phonon-mediated s-wave superconductivity, it
is often difficult to analyze the results of measurements of various physical
quantities because there exist two gaps of different magnitude associated 
with different bands. Their ratio is estimated as 
$r=\Delta^S_0/\Delta^L_0\sim 0.3 - 0.4$ where the larger gap $\Delta^L_0$ is
associated with the two-dimensional $\sigma$-bands and the smaller gap 
$\Delta^S_0$ with the three-dimensional $\pi$-bands. Evidence for two gaps
is provided by the rapid rise of the specific heat coefficient $\gamma_s(H)$
at very low fields.\cite{Yang,Bouquet} Recent measurements of the in-plane
thermal conductivity $\kappa_s(H)$ for fields both parallel and perpendicular
to the c-axis provide more evidence for the existence of two gaps because 
they show a very unusual field dependence at low temperatures. \cite{Solog}
For increasing field, $\kappa_s(H)$ drops rapidly and reaches a minimum at a 
relatively low field, then exhibits an S-shape behavior where a steep rise in
the low-field region is followed by a downward curvature up to about 
$H_{c2}/2$  and finally an upward curvature near $H_{c2}$ for 
${\bf H}\parallel{\bf c}$. The initial drop is attributed to the decrease of the
phononic contribution $\kappa_{ph}(H)$ which is caused by the rapid
enhancement of scattering of phonons by the quasiparticles in the vortex 
cores associated with the $\pi$-band. The steep increase in the low field
region is attributed to the rapid rise of the electronic contribution 
$\kappa_e(H)$ which arises from the the rapid release of mobile 
quasiparticles in the vortex lattice asociated with the $\pi$-band. The final
upward curvature of $\kappa(H)$ near $H_{c2}$ for ${\bf H}\parallel{\bf c}$ is
presumed to be due to the mobile carriers of the vortex lattice associated
with the $\sigma$-band.

     The field dependence of $\gamma_s(H)$ and $\kappa_{es}(H)$ can be
explained qualitatively in terms of a two-band model with two energy 
gaps of different magnitude in the two bands where the band with strong
pairing (the L, or $\sigma$-band) is responsible for the superconductivity 
and the superconductivity in the second band (the S, or $\pi$-band) is
induced by Cooper pair tunneling. \cite{Nakai} The vortices in the S-band
have large vortex cores of radius $\xi^S_0$ and already start to overlap 
in weak fields. This leads to a strong suppression of the small gap at a 
"virtual" upper critical field $H^S_{c2}\sim 3-4$ kOe which is much 
smaller than $H^{(c)}_{c2}\simeq35$ kOe. The large radius of the vortex
core and the field-induced suppression of the smaller gap is consistent
with recent scanning tunneling spectroscopy (STS) measurements. 
\cite{Esk} 

       Theories of the electronic thermal conductivity $\kappa_e$ 
\cite{TF1} and the phononic thermal conductivity $\kappa_{ph}$ 
\cite{TF2} in the mixed state have been developed on the basis of the
Brandt-Pesch-Tewordt (BPT) \cite{BPT,Brandt,PeschThesis} and the 
equivalent Pesch (P) \cite{Pesch1} approximation schemes. Linear
response equations for the vortex state \cite{Klimesch} yield
another expression for $\kappa_e$ which has been applied to 
d-wave pairing superconductivity in the cuprates. \cite{Vekhter}
In the BPT-approximation the normal and anomalous Green's functions
$G$ and $F$ are derived from Gorkov's integral equations with kernel
$\Delta(r_1)G^0_{-\omega}(r_1-r_2)\Delta^{\ast}(r_2)\exp(-i\int^2_1 2eA 
\cdot ds)$ where $\Delta(r)$ is Abrikosov's vortex lattice order 
parameter and $G^0_{-\omega}$ is the quasihole propagator. Thus
these Green's functions take into account the effects of supercurrent
flow and Andreev scattering in the vortex cores as well as quasiparticle
transfer between the vortices. The expression for $\kappa_e$ 
\cite{TF1}, which has been derived from the Kubo formula in analogy 
to the procedure in Ref.~\onlinecite{Amb}, consists essentially of the
$\omega$-integral with the factor 
$(\omega/T)^2\mbox{sech}^2(\omega/2T)$ determining the temperature
dependence, the "destructive" coherence factor arising from the 
combination $(GG^{\dagger}-FF^{\dagger})$, and the total scattering rate
in the denominator consisting of the sum of the quasiparticle 
scattering rates $\gamma$ and $\gamma_A(\omega,H)$ due to 
impurity and Andreev scattering. In the present paper we shall also
calculate the scattering rate $1/\tau_{es}$ due to thermal scattering in
the vortex state, but this becomes comparable to $\gamma$ only
at higher temperatures. The expression for $\kappa_{ph}$ contains,
in addition to the sum of scattering rates of phonons by different
defects, the scattering rate $1/\tau_{ps}$ for phonons scattered by
quasiparticles in the vortex state. The latter expression was
obtained by replacing in the BCS-expression \cite{BRT} the density
of states (DOS) and "coherence" functions by the expressions
resulting from the P-approximation for the vortex state. \cite{TF2}

     The results for $\gamma_s(H)/\gamma_n\,$, 
$\kappa_{es}(H)/\kappa_{en}\,$, and $\tau_{pn}/\tau_{ps}\,$, which were
obtained in Ref.~\onlinecite{TF2} for a single cylindrical or spherical
band and an isotropic s-wave pairing gap, turned out to depend 
sensitively on the reduced impurity scattering rate 
$\delta = \Gamma/\Delta_0 = (\pi/2)\xi_0/\ell$ where $\xi_0$ is the
zero-field coherence length and $\ell$ is the mean free path. If one
takes for the $\pi$-band in MgB$_2\,$ $\xi_{\pi}\sim500\mbox{\AA}$ 
from the STS measurements \cite{Esk} and a mean free path 
$\ell_{\pi} \sim 500-800 \mbox{\AA}\,$, 
one arrives at a reduced sacttering rate
$\delta_{\pi} \sim 1$ which is much larger than the largest value
of $\delta = 0.5$ taken in Ref.~\onlinecite{TF2}. For such a large
impurity scattering rate it is necessary to renormalize the impurity
scattering self energy and to take into account the corresponding
vertex corrections. The vertex corrections yield a renormalization
of the gap function \cite{Brandt,Pesch1} which, near $H_{c2}\,$, 
tends to the renormalization function which has been derived in
the calculation of $H_{c2}\,$. \cite{Helfand} It turns out that this
gap renormalization substantially reduces the dependence of
the functions $\gamma_s(H)/\gamma_n\,$, 
$\kappa_{es}(H)/\kappa_{en}\,$, and $\tau_{pn}/\tau_{ps}(H)$
on impurity scattering.

      Recently it has been shown that 
numerical solutions of the Eilenberger equations yield a spatial
average of the density of states (DOS) which is approximated
very well by the DOS obtained with the BPT- and P-approximations
over the whole field range between $H_{c2}$ and $H_{c1}\,$. 
\cite{Dahm} This result motivates us to calculate the field dependence
of the central parameter of the theory, $\tilde{\Delta}=\Delta\Lambda/v\,$,
from the gap equation in the P-approximation \cite{Dahm} by including
the impurity scattering vertex corrections in analogy to those in the gap
equation of the BPT-approximation. \cite{Brandt} Here $\Delta^2$ is the
spatial average of $|\Delta(r)|^2$ where $\Delta(r)$ is Abrikosov's vortex
lattice order parameter, $\Lambda=(2eH)^{-1/2}=(\pi/2)^{1/2}a$ where $a$
is the lattice constant of the vortex lattice, and $v$ is the Fermi velocity
perpendicular to the uniform field {\bf H}. It turns out that one can fit 
 the exact results of these calculations very well with the relation 
between $\tilde{\Delta}^2$ and $H/H_{c2}$ given by the 
Ginzburg-Landau relationship if one uses an appropriate value of
the Abrikosov parameter $\beta_A = <|\Delta|^4>/(<|\Delta|^2>)^2\,$. 

      We have also investigated the temperature dependence of the
phonon relaxation time ratio $\tau_{pn}/\tau_{ps}(H)$ and
find that this ratio increases rapidly to one with
increasing $H$ in proportion to $[\gamma_s(H)/\gamma_n]^2$ for
temperatures $T < (1/2)\Delta(T,H)\,$. This yields a rapid decrease
of $\kappa_{ph}$ for small fields and relatively low temperatures
if one takes for $\Delta$ the small gap $\Delta^S(T,H)$ of the
$\pi$-band. In this way one can explain the
observed drop of $\kappa$ for small fields and temperatures
near $T_c/6$ which is attributed to the drop of $\kappa_{ph}\,$.
\cite{Solog}

     Since the interband impurity scattering between the $\sigma$- 
and $\pi$-band is small, \cite{Mazin} one can add, in a crude
approximation,  our results obtained for a single isotropic gap
in a cylindrical or spherical Fermi surface (FS) band by weighting
them with the corresponding density of states. If one takes
into account that the "virtual" upper critical field for the
three-dimensional $\pi$-band is much smaller than the upper
critical field of the two-dimensional $\sigma$-band,\cite{Bouquet}
one can explain qualitatively the measured field dependence of 
the specific heat and thermal conductivity in MgB$_2\,$.

\section{Theory of the specific heat and electronic and
phononic thermal conductivity in Abrikosov's vortex
lattice}

     The basis of our theory is the approximation of Brandt, Pesch, 
and Tewordt (BPT) \cite{BPT,Brandt,PeschThesis} for the normal and 
anomalous Green's functions $G$ and $F$ in the mixed state.  A 
simplified version of this theory has been derived by Pesch 
(P) \cite{Pesch1} from the quasiclassical Eilenberger
equations. In this version the $\varepsilon$-integrals of the spatial 
averages of the spectral functions, $-\mbox{Im}G/\pi$ and 
$-\mbox{Im}F/\pi\,$, denoted by $A$ and $B\,$, take on the following
form: \cite{TF2}
\begin{equation}
A(\Omega;\tilde{\Delta},\theta) \equiv  N(\omega,\theta)/N_0= 
\mbox{Re}\left[ 1+ \frac{8\tilde{\Delta}^2_r}{\sin^2\theta}
\left[ 1 + i\sqrt{\pi}\,z\,w(z) \right] \right]^{-1/2}\, ,
\label{A}
\end{equation}
\begin{equation}
B(\Omega;\tilde{\Delta},\theta) = \mbox{Re} \left[
\frac{-i\,\sqrt{\pi}\,2\,(\tilde{\Delta}_r/\sin\theta)\,w(z)}
{ \left\{ 1 + (8\tilde{\Delta}^2_r/\sin^2\theta)[1+i\sqrt{\pi}\,z\, w(z)]
\right\}^{1/2} } \right]\, ,
\label{B}
\end{equation}
\begin{equation}
z = 2[\Omega\,\tilde{\Delta} + i(\Lambda/v)\,\gamma]/\sin\theta\, ;
\qquad \theta = \angle({\bf p},{\bf H})\, ,
\label{z}
\end{equation}
\begin{equation}
\Lambda=(2eH)^{-1/2}\, ,\,\, \tilde{\Delta}= \Delta\Lambda/v\, ,
\,\, \Omega=\omega/\Delta\, ,
\label{Lam}
\end{equation}
\begin{equation}
\tilde{\Delta}^2=(H_{c2}-H)/6\beta_A H\, ,
\quad \Delta^2=\Delta^2_0(T)\left[ 1 - (H/H_{c2})\right]\, ,
\label{Del}
\end{equation}
\begin{equation}
w(z)=\exp(-z^2) {\mathrm{erfc}}(-iz) \, .
\label{w}
\end{equation}
\begin{equation}
\tilde{\Delta}_r=\tilde{\Delta}/D\, ,
\quad D(\Omega)=
1-\int^{\pi/2}_0 d\theta\,2(\Lambda\gamma/v) \sqrt{\pi}\,w(z)\, .
\label{D}
\end{equation}

Here $\Delta_0(T)$ is the s-wave gap in zero field, $H$ is the spatial 
average of the magnetic field, $\Delta^2 = <|\Delta|^2>\,$, where 
$\Delta(r)$ is Abrikosov's vortex lattice order parameter, and $\beta_A$
is the Abrikosov parameter. For the total impurity scattering rate 
$\gamma$ we employ the Born approximation, 
$\gamma = \Gamma \bar{A}$ where $\Gamma = 1/2\tau_n$ and 
$\bar{A}$ is the average of $A$ over the Fermi surface which is 
calculated self-consistently. For a spherical Fermi surface (FS), 
$v\sin\theta$ is the component of the Fermi velocity ${\bf v}({\bf p})$
perpendicular to the field ${\bf H}\,$. For a cylindrical FS parallel to
${\bf c}$ and field ${\bf H}\,||\,{\bf c}\,$, one has to take $\theta = \pi/2$
and $v$ equal to the Fermi velocity in the ab-plane. For a general FS,
$v\sin\theta$ is replaced by the component $v_{\perp}({\bf p})$
perpendicular to the field. The gap renormalization function $D$ in
Eq.(\ref{D}) was first calculated by U. Brandt \cite{Brandt} from the
ladder summation of impurity interaction lines bridging the vertices
at $\Delta$ and $\Delta^{\ast}$ occuring in the self energy part
$G^0_{\omega}\Delta G^0_{-\omega}\Delta^{\ast}G^0_{\omega}$
in Gorkov's integral equation for $G\,$. In the P-approxiamtion 
scheme $D$ takes the form given in Eq.(\ref{D}) which,
near $H_{c2}\,$, tends to the function introduced in Ref.~\onlinecite{Helfand}.

       The field dependence of the DOS, A, and the "coherence factor" 
function, B, are governed by the quantity 
$\tilde{\Delta}=\Delta\Lambda/v = (2\pi)^{-1/2}a/\xi$ where 
$a = (2/\pi)^{1/2}\Lambda$ is the lattice constant of the Abrikosov  
vortex lattice and $\xi = v/\Delta\pi$ is the effective coherence 
length in the plane perpendicular to {\bf H\,}. For 
$H \rightarrow H_{c2}$ and $H\rightarrow 0$ one obtains the limiting
values $\tilde{\Delta}\rightarrow 0$ and 
$\tilde{\Delta}\rightarrow \infty\,$, respectively. The relationships 
between $\tilde{\Delta}$ and $H/H_{c2}$ and between $\Delta$ and 
$H/H_{c2}$ given in Eq.(\ref{Del}) have been derived from the 
GL-theory which is valid near $H_{c2}\,$. To get the general
relationship at lower fields we make use of the gap equation in 
the P-approximation.\cite{Dahm} At $T=0$ this equation can be 
rewritten in the following form:
\begin{equation}
\int^{\infty}_0d\Omega\,[B(\Omega)-C(\Omega)]=0\, .
\label{gapeq}
\end{equation}
Here $B(\Omega)$ is the spectral function of $F$ given in 
Eq. (\ref{B}) and $C(\Omega)$ is obtained by taking the limit
of $B$ for $H\rightarrow H_{c2}$ where 
$\Delta\rightarrow0\,$:
\begin{equation}
C(\Omega)=\mbox{Re}\left[ -i\sqrt{\pi}2(H/H_{c2})^{1/2}
(\tilde{\Delta}_r/\sin\theta) w\left[(H/H_{c2})^{1/2}z\right] \right]
 \Big| _{\Lambda(H_{c2})} \, .
\label{C}
\end{equation}
The gap equation Eq. (\ref{gapeq}) in the P-approximation corresponds
to the gap equation in the BPT-approximation \cite{Brandt} with the 
function $B$ replaced by the spectral function of $F$ derived for the
theory of NMR. \cite{PeschThesis}
Thus Eq. (\ref{gapeq}) is an implicit equation for $H/H_{c2}$
as a function of $\tilde{\Delta}$ for given impurity scattering rate
$\delta= \Gamma/\Delta_0\,$. This calculation is rather complicated
because the impurity scattering self energy 
$\gamma=\Gamma\bar{A}(\Omega)$ occuring in the expression for
$B(\Omega)$ in Eq. (\ref{B}) has to be calculated self-consistently for
each value of $\Omega$ from the expression for  $A(\Omega)$ 
(see Eq. (\ref{A})). For ${\bf H}\,||\,{\bf c}$ and a 
cylindrical FS one has to set $\theta = \pi/2$ in Eq.(\ref{gapeq}), 
and for a spherical FS Eq.(\ref{gapeq}) is averaged over the polar
angle $\theta\,$. In Fig. 1 we show our results for $H/H_{c2}$
versus $\tilde{\Delta}$ obtained from Eq.(\ref{gapeq}) for 
$\theta = \pi/2$ and for $\delta$ = 0.1, 0.5, and 1.0.
These results can be fitted very well by the relationship 
$H/H_{c2}=(1+6\beta_A\tilde{\Delta}^2)^{-1}$ [see Eq.(\ref{Del})]
over the whole field range if one takes $\beta_A$ = 1.1, 1.4, and 1.7
for $\delta$ = 0.1, 0.5, and 1.0. Similar results are obtained for a 
spherical FS if one averages Eq.(\ref{gapeq}) over the polar angle 
$\theta\,$ (see the dashed curves in Fig. 1 for $\delta$ = 0.1, 0.5, 
and 0.8 with $\beta_A$ = 2.3, 2.8, and 3.2). 
 
        In Fig. 2 we have plotted our numerical results for the
normalized DOS at zero energy, 
$A(\Omega=0)=\gamma_s(H)/\gamma_n\,$, versus $H/H_{c2}$
where $\gamma$ is the specific heat coefficient. Here we have
approximated $\tilde{\Delta}$ occuring in Eq. (\ref{A}) by the
analytical expression in Eq. (\ref{Del}) with $\beta_A$ = 1.1, 
1.4, and 1.7 for bare impurity scattering rates 
$\delta=\Gamma/\Delta_0=\,$0.1, 0.5, and 1.0 (solid curves, from
top to bottom). Here we have employed the Born approximation
for the total impurity scattering self energy, 
$\gamma=\Gamma\,\overline{A(0)}\,$, where $\overline{A(0)}$ is 
calculated self-consistently. This means that $\delta$ is 
replaced by $\delta\,\overline{A(0)}\,$. For $\delta=0.1$ and 
$\theta=\pi/2$ the function $A(0)$ is close to the clean limit 
result shown in Ref.~\onlinecite{Dahm} for the s-wave pairing 
state. The latter function has been shown to be very close to the 
numerical solution of the Eilenberger equations over the whole 
field range from $H_{c2}$ to $H=0$ corresponding to $H_{c1}\,$. 
One can see that all curves in Fig. 2 tend to zero in the limit
$H/H_{c2}\rightarrow0\,$,   which is due to the gap
renormalization. For a spherical FS one has to average 
Eq.(\ref{A}) over the polar angle $\theta\,$. For the plot of $A$
versus $H/H_{c2}$ one needs Eq.(\ref{Del}) where $\beta_A$ 
is calculated from the $\theta$-angle average of Eq.(\ref{gapeq}).
The results are shown by the dashed curves in Fig. 2 for
$\delta$ = 0.1, 0.5, and 0.8 with the $\beta_A$-values used in
Fig. 1. It is interesting that the curve for $\delta$ = 0.8 is almost
linear over the whole field region.

     We turn now to the theory of the electronic thermal
conductivity $\kappa_e$ in the vortex state which has been
developed in Ref.~\onlinecite{TF1} by replacing the zero-field
Green's functions $G$ and $F$ in the theory of 
Ref.~\onlinecite{Amb} by the Green's functions of the
BPT-approximation. This theory has been applied to
superconducting states with nodes in the gap such as 
Sr$_2$RuO$_4\,$, the cuprates, and UPt$_3\,$. To save space 
we omit here the expression for $\kappa_e\,$. The main
features of this expression are the following. Most important is the
term Im$\,\varepsilon_0$ in the denominator of the $\omega$-integral
with the well-known factor $\omega^2\mbox{sech}^2(\omega/2T)\,$. This
term Im$\,\varepsilon_0$ corresponds to the scattering rate of
quasiparticles due to impurity and Andreev scattering. The other term
corresponds to the coherence factor of BCS theory. In the zero field
limit both terms in the expression of Ref.~\onlinecite{TF1} tend
correctly to those occuring in the expression of Ref.~\onlinecite{Amb}.
The physical meaning of Im$\,\varepsilon_0$ is the following. The 
equation for the position, $\varepsilon_0\,$, of the pole of the Green's
function $G$ in the BPT-approximation \cite{BPT} yields
Im$\,\varepsilon_0 = \gamma + \gamma_A$ where $\gamma$ is the
total impurity scattering rate and $\gamma_A$ is the imaginary part of
the quasiparticle self energy $\Sigma_{\omega}$ at $\varepsilon_0\,$.
From the kernel of Gorkov's integral equation for $G$ one sees that,
in the spatial representation, 
$\Sigma_{\omega}(r_1,r_2) =-V(r_1,r_2)G^0_{-\omega}(r_1-r_2)\,$, where
$V(r_1,r_2)=\Delta(r_1)\Delta^{\ast}(r_2)\exp(-\imath\int^2_1 2eA \cdot ds)$
and $\Delta(r)$ is Abrikosov's vortex lattice order parameter. From this 
expression one recognizes that $\gamma_A = -\mbox{Im}\Sigma_{\omega}$
is the scattering rate for converting a quasiparticle into a quasihole by
Andreev reflection at $\Delta^{\ast}(r_2)$ and then back into a quasiparticle
at $\Delta(r_1)\,$. The Fourier transform of $\Sigma_{\omega}$ with respect
to the difference coordinate $(r_1-r_2)$ is equal to
\begin{equation}
\Sigma_{\omega}(p)=-\int\,d^3p'\,V(p')G^0_{-\omega}(p-p')\, ,
\label{Sigma}
\end{equation}
with 
\begin{equation}
V(p')=8\pi^2\Delta^2\Lambda^2\delta(p'_z)\exp\left[
%-\Lambda^2\left(p'_x^2+p'_y^2\right)\right]\qquad  % 
-\Lambda^2\left(p_x^{\prime \, 2}+p_y^{\prime \, 2}\right)\right]\qquad
({\bf \hat{z}} \parallel{\bf H})\, .
\label{V}
\end{equation}
For a sperical Fermi surface the result is \cite{BPT}
\begin{equation}
\Sigma_{\omega}(p)=-i\sqrt{\pi}\Delta^2(\Lambda/v\sin\theta)
w[(\omega+i\gamma+\varepsilon)\Lambda/v\sin\theta]
\label{Sigma2}
\end{equation}
where $\varepsilon$ is the normal state energy measured from the Fermi
energy, $\theta=\angle({\bf p},{\bf H})\,$, and $\gamma$ is the total impurity
scattering self energy. 

     In the $\omega\rightarrow0$ limit corresponding to $T\rightarrow0$
one obtains the following explicit expression for Im$\,\varepsilon_0\,$:
\cite{TF1,Brandt}
\begin{eqnarray}
\mbox{Im}\varepsilon_0 &=& \gamma + \gamma_A\, ; \qquad
\gamma_A = (v/\Lambda)\beta \nonumber \\
\beta &=& \tilde{\Delta}^2_r\left\{\left[(\gamma\Lambda/v)^2
+ \tilde{\Delta}^2_r + \eta\sin^2\theta \right]^{1/2} + 
(\gamma\Lambda/v)\right\}^{-1}\, ; 
\quad (\pi^{-1} \le \eta \le 1/2)\, .
\label{Beta}
\end{eqnarray}
All the other quantities $\tilde{\Delta}$ and $\beta$ occuring in the 
expression for $\kappa_e$ (see Eq.(13) of Ref.~\onlinecite{TF1})) are
also renormalized by the replacement 
$\tilde{\Delta}\rightarrow\tilde{\Delta}_r\,$ (see Eq.(\ref{D}) for $\Omega=0$).
Note that $\tilde{\Delta} \sim a/\xi$ and $\Gamma\Lambda/v \sim a/\ell\,$ .
In Fig. 3 we have plotted our results for $\kappa_{es}/\kappa_{en}$ versus
$H/H_{c2}$ in the limit $T \rightarrow 0\,$. The bare impurity scattering rates 
$\delta = \Gamma / \Delta_0$ are 0.1, 0.5, and 1.0 (from bottom to top). For
a spherical FS one has to evaluate the average of $\kappa_e$ over the
polar angle $\theta$ which yields nearly the same results as shown in
Fig. 3 for the plots of $\kappa_{es}/\kappa_{en}$ vs $H/H_{c2}$ if one uses
Eq.(\ref{Del}) with the $\beta_A$ calculated from the average of 
Eq.(\ref{gapeq}) (see the dashed curves in Fig. 3 for $\delta$ = 0.1, 0.5, and
0.8 with the $\beta_A$-values given in Fig. 1). 
 All curves tend to zero in the limit 
$H/H_{c2} \rightarrow 0\,$. One sees that, for
low impurity scattering rates, $\kappa_{es}/\kappa_{en}$ exhibits a strong
upward curvature near $H_{c2}$ which is caused by the rapid reduction
of the Andreev scattering rate $\gamma_A$ with decreasing $\tilde{\Delta}$
or increasing field as can be seen from Eq. (\ref{Beta}). For increasing 
impurity scattering rate $\gamma$ this effect is diminished because 
$\gamma_A$ becomes smaller in comparison to $\gamma$ as can be seen
from the expression in Eq. (\ref{Beta}).

      The main problem in analyzing the measured thermal conductivity
$\kappa$ in the mixed state is to separate the electronic and phononic
contributions $\kappa_e$ and $\kappa_{ph}\,$. \cite{Solog} The general
expression for $\kappa_{ph}$ (see Eq. (4) of  Ref.~\onlinecite{TF2}) 
contains, in addition to other phonon relaxation times due to defect 
scattering, the relaxation time $\tau_p(\omega)$ due to scattering of 
phonons of frequency $\omega$ by quasiparticles which changes
drastically in the BCS state. The expression for $1/\tau_{ps}$ in the
BCS state \cite{BRT} has been extended to the vortex lattice state by
replacing the BCS DOS $E/\varepsilon\,$, where 
$\varepsilon=(E^2 - \Delta^2)^{1/2}\,$, and the function 
$\Delta/\varepsilon$ occuring in the BCS coherence factor, with the
spectral functions $A$ and $B$ of Eqs. (\ref{A}) and (\ref{B}). In the
zero field limit the latter expressions tend correctly to the BCS
functions with $\omega \equiv E\,$. The resulting ratio of phonon
relaxation times in the normal and mixed states is given by
(see Eq. (3) of Ref.~\onlinecite{TF2}):
\begin{eqnarray}
\tau_{pn}/\tau_{ps} = g(\Omega_0) &=& [1-\exp(-\Omega_0/y)]
(2/\Omega_0)\int_0^{\infty}d\Omega\,f[(\Omega-\Omega_0/2)/y]\,
f[-(\Omega+\Omega_0/2)/y] \nonumber \\
& & \times \left[A(\Omega-\Omega_0/2) \, A(\Omega+\Omega_0/2) -
B(\Omega-\Omega_0/2) \, B(\Omega+\Omega_0/2) \right]\, .
\label{tauratio}
\end{eqnarray}
Here, $\Omega_0=\omega_0/\Delta$ is the phonon energy
$\omega_0$ divided by $\Delta=\Delta(T,H)\,$, f is the Fermi
function, and $y=T/\Delta(T,H)\,$. In Ref.~\onlinecite{TF2} we
calculated $g$
in the limit $T\rightarrow0\,$. It was found that in the limit
$\tilde{\Delta}\rightarrow\infty\,$, or $H/H_{c2}\rightarrow 0\,$, 
g tends correctly to the BCS step functon with a step at 
$\Omega_0 = 2\,$, or $\omega_0 = 2\Delta\,$. This step function
is washed out as 
$\tilde{\Delta}\rightarrow 0\,$, or $H/H_{c2}\rightarrow1\,$, and it
tends to the constant, 1. Since it is important to know the temperature
dependence of $\kappa_{ph}\,$, and thus of $\tau_{pn}/\tau_{ps}=g\,$,
for analyzing the measured $\kappa$ as a function of $T$ at fixed
field, \cite{Solog} we have now calculated the $T$-dependence of the
expression in Eq.(\ref{tauratio}). For this purpose we make use of the
fact that, according to the general expression for $\kappa_{ph}$ (see
Eq.(\ref{Lam}) of Ref.~\onlinecite{TF2}), the maximum of the integrand
in that x-integral occurs at about $x_m=\omega_{0m}/T\simeq4\,$.
Therefore it is a good approximation to make the variable 
transformations $\Omega_0 = x_my$ and $\Omega = ty$
in Eq.(\ref{tauratio}). This yields
$\tau_{pn}/\tau_{ps}=g(x_my)$ as a function of the $T$-dependent
quantity $y$ for fixed $\tilde{\Delta}\,$, or field $H/H_{c2}\,$. In 
Fig. 4 we have plotted $g(x_my)$ versus $y$ for
constant $x_m=4$ and constant $\tilde{\Delta}=$ 0.2, 0.3, and 0.6,
or $H/H_{c2}=$ 0.78, 0.62, and 0.29 (see Fig. 1) for $\delta=0.1\,$. 
In the limit $H/H_{c2}\rightarrow0\,$, 
one obtains a step function with a step at $y=1/2\,$, that is, 
$\Omega_0 = 4y = 2$ or $\omega_0=2\Delta\,$, as it should be. One
sees, that for intermediate values of the field,  
$\tau_{pn}/\tau_{ps}=g$ considered as a function of $T$ is smeared
out and tends to one for high fields. In the limit 
$y\rightarrow0\,$, or $\Omega_0\rightarrow0$
Eq.(\ref{tauratio}) yields 
$g(\Omega_0=0)=[A(\Omega_0=0,\,\tilde{\Delta}\,,\theta)]^2\,$. 
If the scattering rate $1/\tau_{ps}=(1/\tau_{pn})g$ occuring in the 
denominator of the $\omega$-integral in the expression for 
$\kappa_{ph}$ (see Eq.(4) in Ref.~\onlinecite{TF2}) is comparable to 
or larger than the other scattering rates due to phonon-defect 
scattering, we expect that $\kappa_{ph}$ rapidly decreases
for increasing field for temperatures smaller or close to 
$T\simeq(1/2)\Delta(T,H)\,$. In the measurements of
$\kappa$ one observes a fast drop of $\kappa$ in small fields
at temperatures around $T_c/6\,\,$ \cite{Solog} which can be 
attributed to the fast drop of $\kappa_{ph}\,$ due
to scattering by quasiparticles in the $\pi$-band vortex lattice.

     It is interesting to compare the temperature dependence of the
phonon lifetime $\tau_p$ with the lifetime $\tau_e$ of a 
quasiparticle due to scattering by acoustic phonons. This 
scattering rate has been derived for the BCS state in 
Ref.~\onlinecite{Tew} (see Eq.(3.9)). We extend this expression
for the scattering rate $1/\tau_{es}$ in anology to our procedure
for $1/\tau_p$ from the BCS to the vortex state by replacing the
density of states by the DOS $A(\Omega)$ in Eq.(\ref{A}) and the
coherence factor terms by $B(\Omega)$ in Eq.(\ref{B}). Then we
obtain:
\begin{eqnarray}
1/\tau_{es}(\omega) 
&=& 
C[f(-\beta\omega)]^{-1} A^{-1}(\Omega)
\int^{\infty}_0 d\omega_0 \,\omega^2_0 \left\{ 
A(\Omega) \, A(\Omega-\Omega_0) -
B(\Omega) \, B(\Omega-\Omega_0)  \right. \nonumber \\
& & \times  f[-\beta(\omega-\omega_0)] [1+b(\beta\omega_0)] \nonumber \\
& &+ \left. [A(\Omega) \, A(\Omega+\Omega_0) - B(\Omega) \, B(\Omega
+\Omega_0)]
f[-\beta(\omega+\omega_0)] b(\beta\omega_0)  \right\}\, .
\label{taues}
\end{eqnarray}
Here $C$ is a constant arising from the electron-phonon matrix elements, 
$\beta=1/T\,$, $b(x)=(e^x-1)^{-1}\,$, and $\omega_0$ is the phonon frequency.
We use this expression now in the calculation of the electronic thermal
conductivity $\kappa_e\,$. We mentioned above that the integrand of the
$\omega$-integral in the expression for $\kappa_e$ given in Ref.~\onlinecite{TF1}
contains the term Im$\varepsilon_0=\gamma + \gamma_A$ in the denominator
which corresponds to the sum of relaxation rates of a quasiparticle due to 
impurity and Andreev scattering. The effect of thermal scattering by phonons
can now be taken into account by adding $1/\tau_{es}$ to Im$\varepsilon_0\,$.
The temperature dependence of $\kappa_e$ is governed by the factor 
$(\omega/T)^2\mbox{sech}^2(\omega/2T)$ in the integrand of the $\omega$-integral 
which yields the main contribution of the integrand from the vicinity of
$x=\omega/T$ at about $x_m=2\,$. Therefore we can estimate the temperature
dependence of $1/\tau_{es}$ by carrying out the following variable transformations
in Eq.(\ref{taues}): $x=\omega/T\,$, $t=\omega_0/T\,$, $\Omega=x\,y\,$, and
$\Omega_0=t\,y$ with $y=T/\Delta(T,H)\,$. This yields $1/\tau_{es}$ as a function
of $\Omega=x\,y$ and $x\,$. The normal state relaxation rate $1/\tau_{en}$ is
obtained from Eq.(\ref{taues}) by setting $A=1$ and $B=0\,$. In Fig. 5 we have 
plotted $\tau_{en}/\tau_{es}$ as a function of $y$ for constant
$x=x_m=2$ and $\tilde{\Delta}=\,$ 0.2, 0.3, and 0.6 ($H/H_{c2}$= 0.78, 0.62, and 
0.29, for $\delta=0.1\,$, see Fig. 1). One sees that for increasing $\tilde{\Delta}\,$,
or decreasing field, this function tends to a step function where the step occurs at
$y=1/2\,$, that is, at a quasiparticle energy $\Omega=\omega/\Delta=1\,$. This 
result agrees with the previous result for the BCS state. \cite{Tew} One recognizes
that the variations of $\tau_{en}/\tau_{es}$ as a function of $y$ are much smaller
than those of the phonon lifetime ratio $\tau_{pn}/\tau_{ps}$ shown in Fig. 4 
where, for increasing $\tilde{\Delta}\,$, a step occurs at about 
$y=\Omega/4=1/2\,$, i.e., at a phonon energy $\Omega_0=\omega_0/\Delta =2\,$.
The contribution of $1/\tau_{es}$ to the relaxation rate Im$\varepsilon_0$ 
occuring in the electronic thermal conductivity $\kappa_e$ increases with
temperature like $T^3$ because $1/\tau_{en}$ is proportional to $T^3\,$.
In the presence of moderate impurity scattering this becomes comparable
with $\gamma$ only at temperatures such that $y=T/\Delta(T,H)>1$ where,
according to Fig. 5, the ratio $\tau_{en}/\tau_{es}$ is nearly one. The 
expression for $1/\tau_{es}$ in Eq.(\ref{taues}) has been calculated in
analogy to Ref.~\onlinecite{Tew} for quasiparticle scattering by acoustic 
phonons. Instead of this spectrum, $C\omega_0^2\,$, one should employ in
the expression for $1/\tau_{es}$ the transport Eliashberg functions 
$\alpha^2_{tr}F_{nn}(\omega_0)\,$, where $n=\sigma,\,\pi\,$, which have been
calculated for MgB$_2\,$. \cite{Kong} Since the coupling constant 
$\lambda_{\sigma\sigma}=0.81 > \lambda_{\pi\pi}=0.41\,$, we conclude that 
for the $\sigma$-band the field dependence of $1/\tau_{es}$ should be
taken into account while, for the $\pi$-band, $1/\tau_{es}$ can safely be
approximated by $1/\tau_{en}\,$.

\section{Conclusions}

     We have calculated the density of states $A(\omega)\,$, the electronic
thermal conductivity $\kappa_e\,$, and the scattering rates 
$1/\tau_p(\omega)$ of a phonon and $1/\tau_e(\omega)$ of a quasiparticle
for a single isotropic gap over the entire field range of the mixed state.
That the BPT\, \cite{BPT,Brandt,PeschThesis} and P \, \cite{Pesch1} 
methods for these calculations are valid approximation schemes has been
proved by comparison of the results for $A$ with the solutions of the
Eilenberger equations for Abrikosov's vortex lattice order parameter.
\cite{Dahm} By solving the gap equation including impurity self energy and
vertex corrections (see Fig. 1), we have shown that the relation between the
central parameter of the theory, $\tilde{\Delta}=\Delta\,\Lambda/v\,$, and
$H/H_{c2}$ can be very well approximated by the Ginzburg-Landau (GL)
relation [see Eq.(5)] over the entire field range if one uses an appropriate
value of the Abrikosov parameter $\beta_A\,$. This result is connected with
the fact that the Abrikosov order parameter is a solution of the linearized GL
equations which allows extension of the BPT method over the entire region
of linear magnetization. \cite{EHBrandt} 

     The second main result of the present paper is that the vertex corrections
due to impurity scattering \cite{Brandt,Pesch1} substantially reduce the
dependence of the functions $\gamma_s(H)/\gamma_n\,$, 
$\kappa_{es}(H)/\kappa_{en}\,$, and $\tau_{pn}/\tau_{ps}(H)$ on the amount
of impurity scattering. The impurity scattering rate 
$\delta=\Gamma/\Delta_0=(\pi/2)(\xi_0/\ell)$ is renormalized here in the Born
approximation $\Gamma \rightarrow \gamma=\Gamma\,\bar{A}(\omega)$ where
$\bar{A}(\omega)$ is calculated self-consistently from Eq.(\ref{A}). The vertex
corrections yield a renormalization of the gap $\tilde{\Delta}_r$ due to the
function $D$ [see Eq.(\ref{D})]. As can be seen from Fig. 2, the functions 
$\gamma_s(H)/\gamma_n$ for $\delta=$0.1, 0.5, and 1.0 are not very
different from each other. The results are however quite different if we set
the renormalization function $D$ equal to one. Then the slope of 
$\gamma_s(H)/\gamma_n$ versus $H/H_{c2}$ at $H=0$ rises rapidly with
$\delta$ and becomes very large for $\delta = 1\,$. It should be noticed that
all curves in Fig. 2 tend to zero in the zero field limit as they should to be 
in accordance with Anderson's theorem for "dirty" superconductors. Our
result that the effect of impurity scattering on the field dependence of
$\gamma_s(H)/\gamma_n$ for a single isotropic gap is rather small and 
does not yield a rapid rise for small fields, gives support for the two-gap
model in interpreting the measurements of $\gamma_s(H)/\gamma_n$ 
on single crystals of MgB$_2\,$. \cite{Bouquet} According to this model, the
first steep increase of $\gamma_s(H)/\gamma_n\,$, which is nearly the
same for applied fields parallel and perpendicular to the c-axis, arises from
the behavoir of the small gap associated with the three dimensional 
$\pi$-band. This in effect closes for increasing field due to the overlap of the
vortex cores with large radius $\xi_{\pi}\sim 500 \mbox{\AA}\,$ \cite{Esk} 
at a relatively small "virtual" upper critical field $H^S_{c2} \sim$ 3 - 4 kOe.
The remaining superconducting contribution above $H^S_{c2}$ arises from
the large gap associated with the two dimensional $\sigma$-band which
closes at an upper critical field $H^{(c)}_{c2} \sim$ 32 kOe and
$H^{(ab)}_{c2} \sim$ 160 kOe. The field dependence of the $\pi$-band and
the $\sigma$-band contributions to $\gamma_s(H)/\gamma_n$ is
presumably given by the almost linear curves in Fig. 2 for $\delta=0,8$ and
$\delta=0.5$ with $H_{c2}$ equal to $H^{S}_{c2}$ and $H^{(c)}_{c2}$ or
$H^{(ab)}_{c2}\,$, respectively. 

      The results for $\kappa_{es}(H)/\kappa_{en}$ shown in Fig. 3 depend on
the impurity scattering rate $\delta$ primarily because the  effect of Andreev
scattering on the upward curvature becomes relatively smaller as $\delta$
increases. Although  the slope at $H=0$ is relatively small even for 
$\delta=1\,$, it becomes very large at  $\delta=1$ if the renormalization
of the gap $\tilde{\Delta}_r = \tilde{\Delta}/D$ due to the vertex correction 
$D$ in $\kappa_e$ [see Eq.(\ref{Beta})] is neglected. This new result has the 
consequence that the observed steep rise of $\kappa_e(H)$ for small fields
parallel or perpendicular to the c-axis in single-crystalline MgB$_2\,$ 
\cite{Solog} can again be explained only in terms of the two-gap model 
where the smaller gap in the three dimensional $\pi$-band closes in effect
at a "virtual" upper critical field $H^{S}_{c2} \sim $3 - 4 kOe due to the
overlap of the vortex cores with large radius $\xi_{\pi}\,$. \cite{Esk} At
higher fields $\kappa_e(H)$ tends to saturate until, in the vicinity of
$H^{(c)}_{c2} \sim $32 kOe, the contribution to $\kappa_e$ from the larger
gap associated with the cylindrical $\sigma$-band exhibits an upward
curvature. The contributions to $\kappa_e$ arising from the $\pi$- and
$\sigma$-bands are presumably given by the curves in Fig. 3 for 
$\delta=0.8$ and $\delta=0.5$ with the corresponding $H_{c2}$ equal to
$H^S_{c2}$ and $H^{(c)}_{c2}$ where the curve for $\delta=0.5$ yields the
measured upward curvature near $H_{c2}\,$.

    The rapid decrease of $\kappa$ observed at low fields, which is
attributed to the decrease of the phononic thermal conductivity 
$\kappa_{ph}\,$, \cite{Solog} can be explained by the field and relevant
temperature dependence of the ratio of phonon lifetimes 
$\tau_{pn}/\tau_{ps}(H) = g(H)$ shown in Fig. 4. One sees that $g(H)$ tends
rapidly to the constant one for increasing field and values of 
$y<1/2$ corresponding to temperatures $T<(1/2)\Delta(T,H)\,$. If we 
assume that $\Delta$ is the smaller gap associated with the $\pi$-band, 
then the $T$ satisfying this condition are much smaller than $T_c\,$.
However, at the lowest temperatures, this effect on $\kappa_{ph}$ is
strongly reduced because the scattering rate $1/\tau_{ps}$ occuring in the
denominator of the $\omega$-integral for $\kappa_{ph}$ is multiplied by
$T/T_c$ in comparison to the scattering rate due to sample boundary 
scattering (see Eq.(4) of Ref.~\onlinecite{TF2}).

      We have calculated $\kappa_e(H)$ in the limit $T\rightarrow0\,$, or
$\omega\rightarrow 0\,$. For finite $T$ one has to solve the 
transcendental equation determining the pole $\varepsilon_0$ of the 
BPT-Green's function \cite{TF1} for each finite value of $\omega$ 
occuring in the $\omega$-integral for $\kappa_e\,$. It should be pointed
out that this calculation is avoided in the expression for $\kappa_e$ 
which was derived from linear response theory. \cite{Klimesch,Vekhter}
For higher temperatures one should add to the sum of scattering rates 
$\gamma$ and $\gamma_A$ due to impurity and Andreev scattering 
[see Eq.(\ref{Beta})], the scattering rate $1/\tau_{es}(H)$ due to 
scattering of quasiparticles by phonons. The field and relevant
temperature dependence of the ratio $\tau_{en}/\tau_{es}(H)$ is shown in 
Fig. 5 as a function of $y=T/\Delta(T,H)\,$. One sees that this ratio is 
nearly equal to one for values of $y>1$ corresponding to temperatures 
$T > \Delta(T,H)\,$. For the small gap associated with the $\pi$-band this
condition is probably satisfied at those temperatures where 
$1/\tau_{en} \sim T^3$ becomes comparable to the intraband impurity 
scattering $\gamma$ for the $\pi$-band. However, for the large gap
associated with the $\sigma$-band, this condition might not be satisfied. 
We conclude from these estimates that, for the $\pi$-band, $1/\tau_{es}$ 
can be approximated by $1/\tau_{en}$ while, for the $\sigma$-band, the
field dependence of $1/\tau_{es}(H)$ should be taken into account. These
considerations are supported by detailed calculations which yield a 
larger intraband impurity scattering rate $\gamma$ for the $\pi$-band
than for the $\sigma$-band, \cite{Mazin} and a larger scattering rate 
$1/\tau_{en}$ for the $\sigma$-band than for the $\pi$-band because the
intraband electron-phonon coupling constant satisfies 
$\lambda_{\sigma\sigma} > \lambda_{\pi\pi}\,$. \cite{Kong}

     Our assumption that we can simply add our results obtained for two
different isotropic gaps and different upper critical fields in the $\pi$- 
and $\sigma$-bands is a crude approximation in view of the actual 
situation in MgB$_2\,$. It is true that the interaction due to interband 
impurity scattering between the $\sigma$- and $\pi$-bands is small. 
\cite{Mazin} However, we have neglected the pairing interaction
between the two bands which presumably leads to Cooper pair 
tunneling from the $\sigma$-band with strong pairing to the $\pi$-band 
where it leads to giant vortex cores and where superconductivity is
maintained above the virtual $H^S_{c2}$ up to $H_{c2}\,$. \cite{Nakai}
Nevertheless we believe that our clear-cut results for the mixed state
of a single isotropic gap are helpful in analyzing the complicated
behavior of different physical quantities in MgB$_2\,$. 

     Another approximation is that we have neglected higher harmonic
oscillator components N of Abrivosov's vortex lattice order parameter
which lead to a distortion of the lattice at lower fields. \cite{Dahm}
For an s-wave superconductor the BPT method breaks down at low
fields where properties are determined by the states bound to the
vortex cores. However, STS measurements in MgB$_2$ show an 
absence of localized states in the cores. \cite{Esk}

\acknowledgements
 
We would like to thank T. Dahm  and K. Scharnberg for valuable
discussions. 
\newpage
\newpage
\vspace{2cm}
\begin{figure}
\centerline{\psfig{file=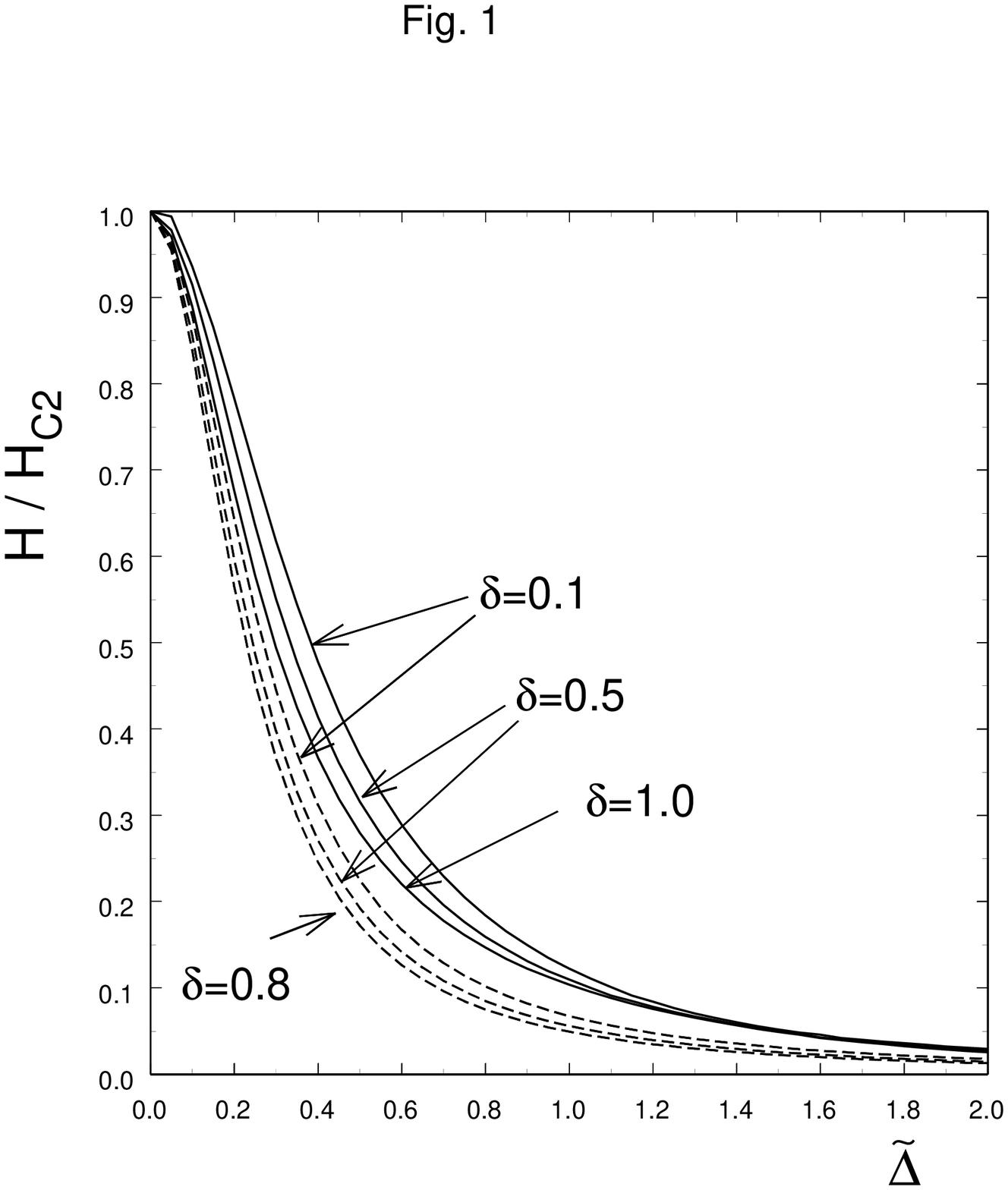,width=18cm,angle=0}}
\vskip -4cm
\caption{Solutions of gap equation for $H/H_{c2}$ vs $\tilde{\Delta}=
\Delta\Lambda/v$ ($\Lambda=(2eH)^{-1/2}$), for impurity
scattering rate $\delta=$ 0.1, 0.5, and 1.0, for $\theta=\pi/2\,$ 
(solid curves). The fits are obtained from the analytical 
expression with the Abrikosov parameter
$\beta_A=$ 1.1, 1.4,  and 1.7 [see Eq.(\ref{Del})]. The dashed
curves show the $\theta$-angle averages for
$\delta=$ 0.1, 0.5, and 0.8 with fit parameter values
$\beta_A=$ 2.3, 2.8,  and 3.2. }
\label{fig1}
\end{figure}
\begin{figure}
\centerline{\psfig{file=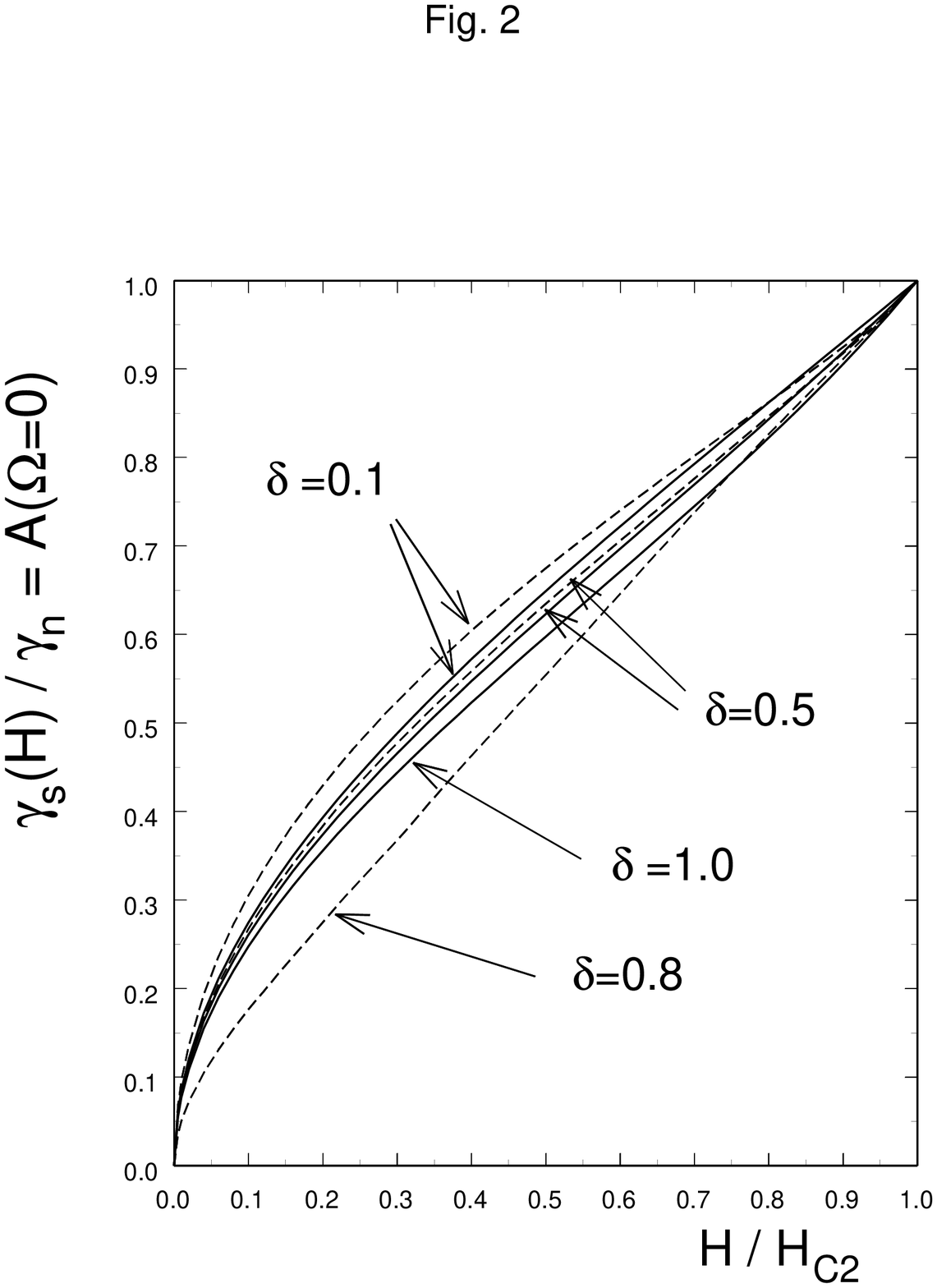,width=18cm,angle=0}}
\vskip -1cm
\caption{Ratio of specific heat coefficients, $\gamma_s/\gamma_n\,$,  vs
$H/H_{c2}$ for reduced impurity scattering rates 
$\delta=\Gamma/\Delta_0$ = 0.1, 0.5, and 1.0 (solid curves)
for $\theta=\pi/2$ (cylindrical Fermi surface(FS)). The dashed
curves show the $\theta$-angle averages (spherical FS) with
$\delta$ and $\beta_A$ as in Fig. 1.}
\label{fig2}
\end{figure}
\begin{figure}
\centerline{\psfig{file=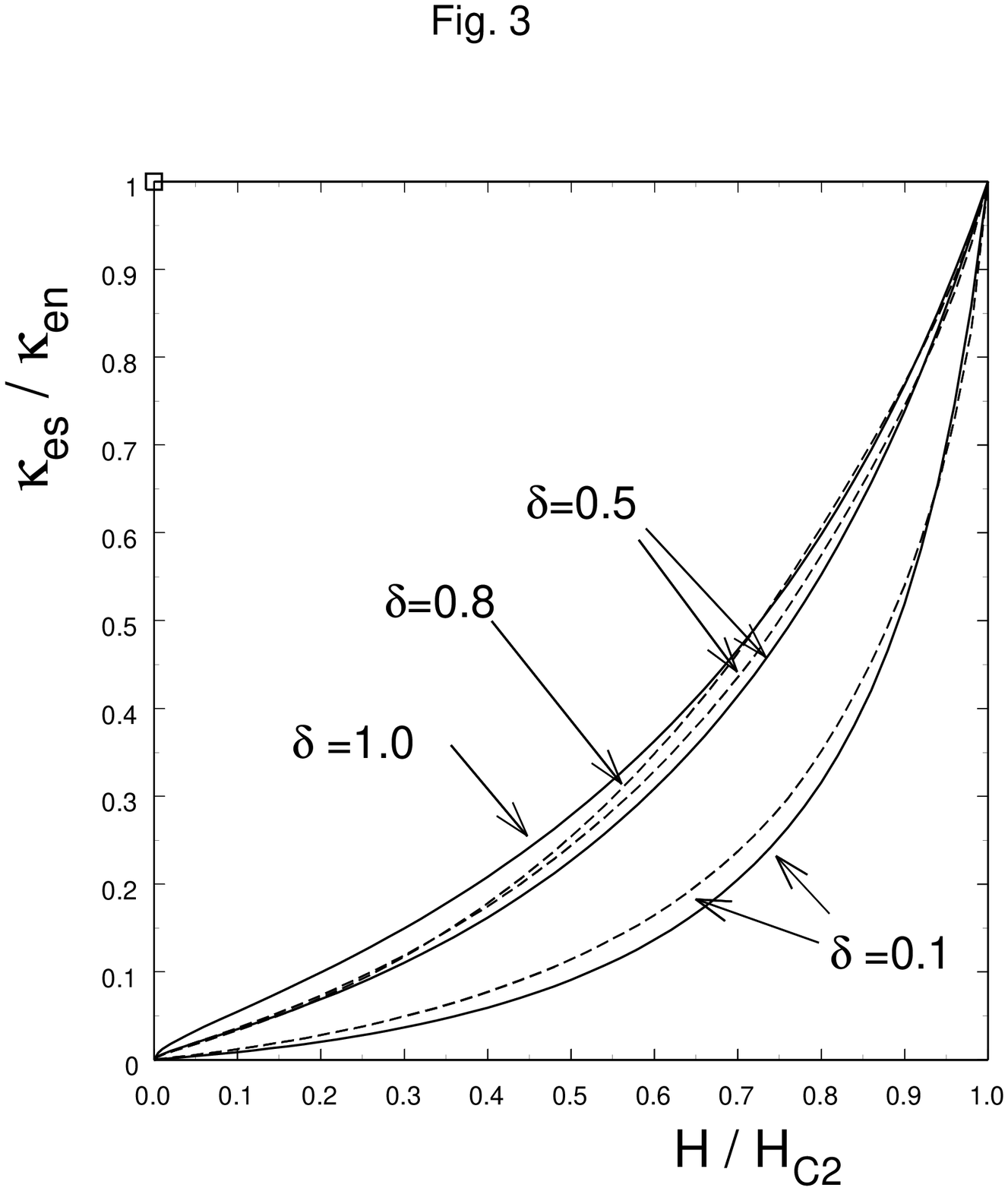,width=18cm,angle=0}}
\vskip -3cm
\caption{Electronic thermal conductivity ratio $\kappa_{es}/\kappa_{en}$ vs 
$H/H_{c2}$ for $\theta=\pi/2$ (solid curves) and for the $\theta$-angle
averages (dashed curves). Notation as in Fig. 2.}
\label{fig3}
\end{figure}
\begin{figure}
\centerline{\psfig{file=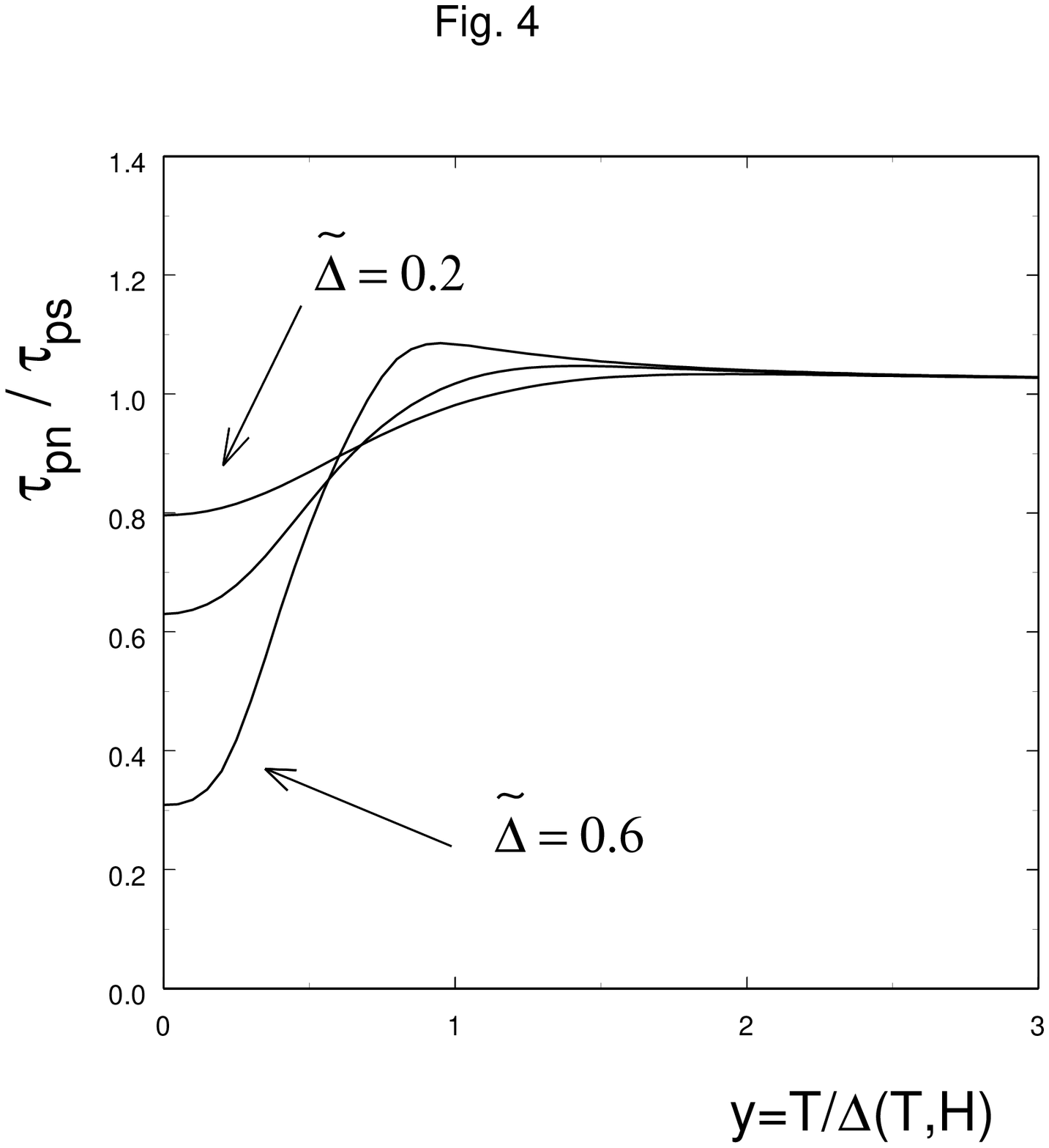,width=18cm,angle=0}}
\vskip -4cm
\caption{Ratio $\tau_{pn}/\tau_{ps}$ of phonon lifetimes vs $y=T/\Delta(T,H)$
corresponding to dominant phonon energy $\omega_0=4\Delta y= 4T$ 
which determines the phononic thermal conductivity $\kappa_{ph}\,$.
The constant fields are $H/H_{c2}=$ 0.29, 0.62, 0.78, or, 
$\tilde{\Delta}=$ 0.6, 0.3, 0.2 (from bottom to top) and $\delta=0.1\,$. }
\label{fig4}
\end{figure}
\begin{figure}
\centerline{\psfig{file=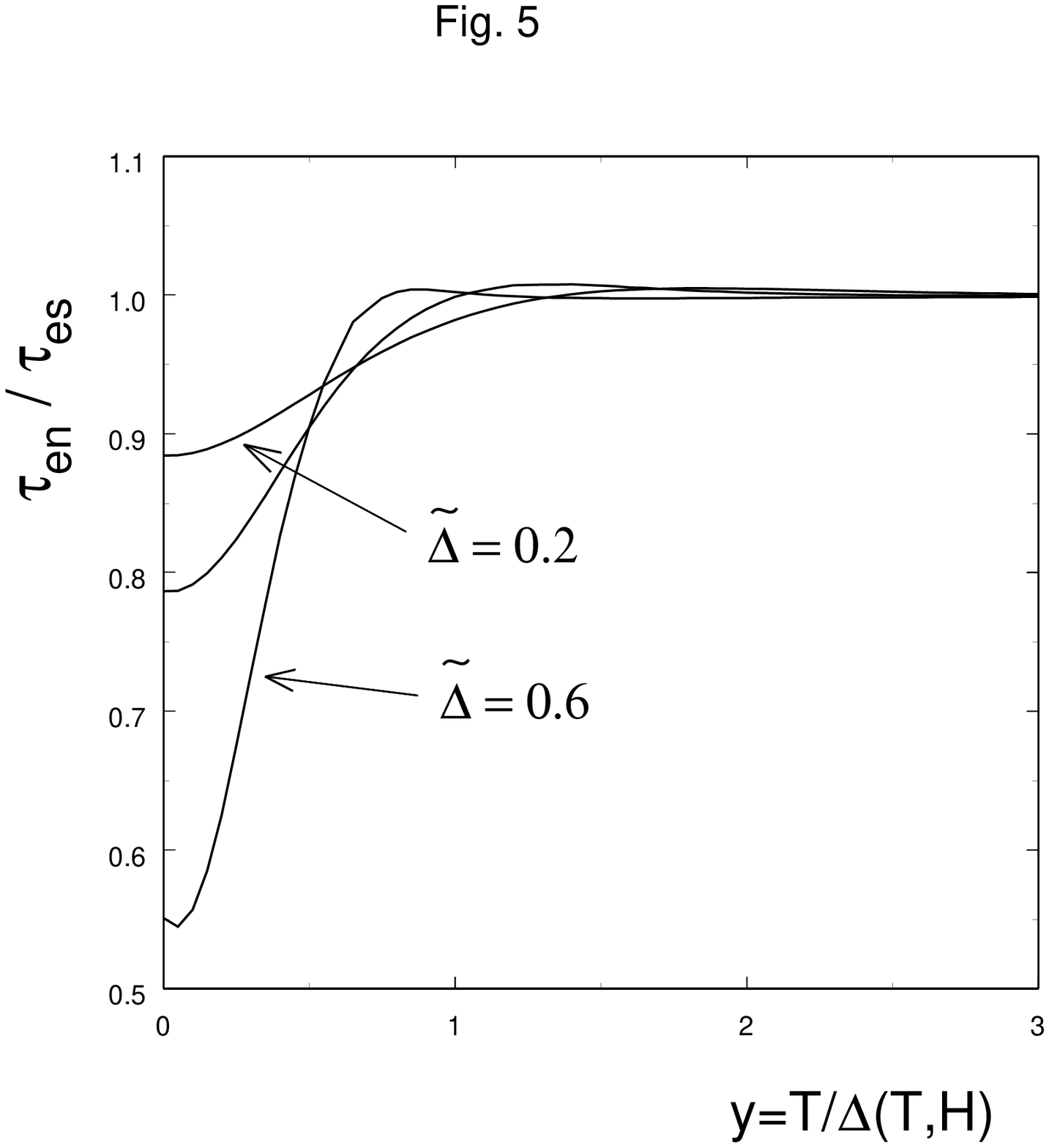,width=18cm,angle=0}}
\vskip -4cm
\caption{Ratio $\tau_{en}/\tau_{es}$ of quasiparticle lifetimes due to phonon
scattering vs $y=T/\Delta(T,H)$ corresponding to dominant quasiparticle
energy $\omega=2\Delta y= 2T$ which determines the electronic thermal
conductivity $\kappa_{e}\,$. The constant fields are 
$H/H_{c2}=$ 0.29, 0.62, 0.78, or, $\tilde{\Delta}=$ 0.6, 0.3, 0.2
(from bottom to top), for $\delta=0.1\,$. }
\label{fig5}
\end{figure}
\end{document}